# Studies of CoFeB crystalline structure grown on PbSnTe topological insulator substrate

A. K. Kaveev[1], N. S. Sokolov[1], S. M. Suturin[1], N. S. Zhiltsov[1], A.E. Klimov[2], O. E. Tereshchenko[2,3]

[1]*Ioffe Institute, 194021 Saint - Petersburg, Russia*
[2] *Rzhanov Institute of Semiconductor Physics, 630090 Novosibirsk, Russia*
[3]*Novosibirsk State University, 630090 Novosibirsk, Russia*

**Abstract:** $Co_{40}Fe_{40}B_{20}$ layers were grown on the $Pb_{0.71}Sn_{0.29}Te$ topological insulator substrates by laser molecular beam epitaxy (LMBE) method, and the growth conditions were studied. The possibility of growing epitaxial layers of a ferromagnet on the surface of a topological insulator was demonstrated for the first time. The $Co_{40}Fe_{40}B_{20}$ layers obtained have a bcc crystal structure with a crystalline (111) plane parallel to the (111) PbSnTe plane. The use of three-dimensional mapping in the reciprocal space of reflection high electron diffraction (RHEED) patterns made it possible to determine the epitaxial relationship of main crystallographic axes between the film and the substrate of topological insulator. Quenching of some reflections in diffraction pattern allows confirmation of the substrate stoichiometry.

## 1. Introduction

Recently, expressed interest in the field of spintronics is related to topological insulators, characterized by a unique electronic structure of surface states [1, 2]. Topological insulators are characterized as conventional insulators in the volume and as objects with Dirac cone of conducting spin-polarized states on the surface. These states, due to the requirement of time-reversal symmetry, are topologically protected from external non-magnetic effects. Topological crystalline insulators based on $Pb_{1-x}Sn_xTe$ solid solution are the extension of 3D topological insulators whose surface states are protected by crystal symmetries, rather than by time-reversal symmetry. The study of the interface of a ferromagnetic -topological insulator is of interest due to the effect of magnetic proximity, acting as a factor breaking the symmetry with respect to inversion of the time of surface topological states [3]. This effect is manifested in the influence of the exchange field of a ferromagnet on topological states. An inverse effect is also interesting to study. It is the effect of the spin-polarized current of the surface states of topological insulators (TI) on the magnetic order in a ferromagnet (FM). The idea of mutual control of the surface states of a topological insulator and the magnetization of a ferromagnet is promising in the field of creating system with the control of spin [4]. In the present work, the studies of the system $Co_{40}Fe_{40}B_{20}$ / $Pb_{0.71}Sn_{0.29}Te$ (hereinafter – CoFeB and PbSnTe, respectively) were carried out. The aim of the work was to obtain mono-crystalline epitaxial layers of CoFeB on the PbSnTe surface.

## 2. Experimental

The CoFeB layers were grown on PbSnTe (111) surface with use of laser MBE system (produced by Surface, GmbH.) based on KrF excimer laser. Crystalline topological insulator $Pb_{1-x}Sn_xTe(111)$ epitaxial films were grown on $BaF_2$ substrate. Clean TI surface was obtained using chemical treatment in HCl-isopropanol solution and subsequent vacuum anneal at 300°C during 30 minutes under $10^{-8}$ mBar pressure. CoFeB layers were grown on PbSnTe at 150°C in argon atmosphere under $2.5\times10^{-3}$ mBar pressure. The thickness of the layers was 10 nm. RHEED analysis was performed *in-situ* with use of built-in diffractometer at 30 kV. We have carried out an image analysis via special software developed in our group [5]. The software allows one to plot 3D projections of RHEED patterns in the reciprocal space for concrete zone axis. The 3D reciprocal space mapping allows us to improve of the conventional RHEED (or LEED) measurement approach. It uses the technique of slicing the reciprocal space with a sequence of Ewald spheres. This is achieved by taking long sequence of electron diffraction patterns while changing in small steps the orientation of the sample (azimuth or tilt) with respect to the incident electron beam. The intensity distribution of the reflections in reciprocal space is found from the combining of acquired images by putting together the spherical cross sections represented by individual patterns.

## 3. Crystalline structure of CoFeB on PbSnTe

Fig. 1 shows RHEED reciprocal space maps of CoFeB layer and PbSnTe substrate with superimposed model reflections. The maps on the left show the reciprocal space cross-sections containing a single reciprocal space zone – with the [111] zone axis for PbSnTe and the [111] zone axis for CoFeB. The maps on the right show the plan-view reciprocal space projections onto the plane parallel to the surface normal. The observed RHEED patterns show pronounced streaks combined with transmission spots which allow supposition of the moderate surface smoothness. The results of reciprocal lattice modeling are superimposed on the maps in Fig. 1 allowing us to conclude that CoFeB grows epitaxially with the body-centered cubic (bcc) crystal structure oriented with its [111] axis perpendicular to the surface, on the face centered cubic (fcc) crystal structure of PbSnTe (described in [6]). The in-plane CoFeB [0-11] axis is oriented along the PbSnTe [-110] axis favored by the fact that interplanar distance 4.49 Å of PbSnTe in [01-1] direction is close to the distance 4.05 Å in [10-1] direction of CoFeB (see right part of Fig. 1). Quenching of some reflections in side view of PbSnTe (for example h 1 l, where h – odd, l - odd) is characteristic for the diffraction patterns of the simple cubic type of unit cell. This fact may be explained by the close form factor values of Sn and Te. As this fact, so relatively small amount of Sn conducts PbSnTe fcc crystal structure very similar to the "simple cubic" from the point of view of the diffraction properties. In case of x=0.5 PbSnTe represents clear fcc structure with two atoms (Pb and Sn, or Pb and Te) in the basis [6]. In case of x=0 it is the simple cubic structure. In case of the intermediate values of x value, some sites are occupied be the one type of atoms (Pb or Sn) in excess to second one. Therefore the crystal represents a "mixture" of fcc and simple cubic structure, closer to the last one depending on how large is the difference of x value from 0.5.

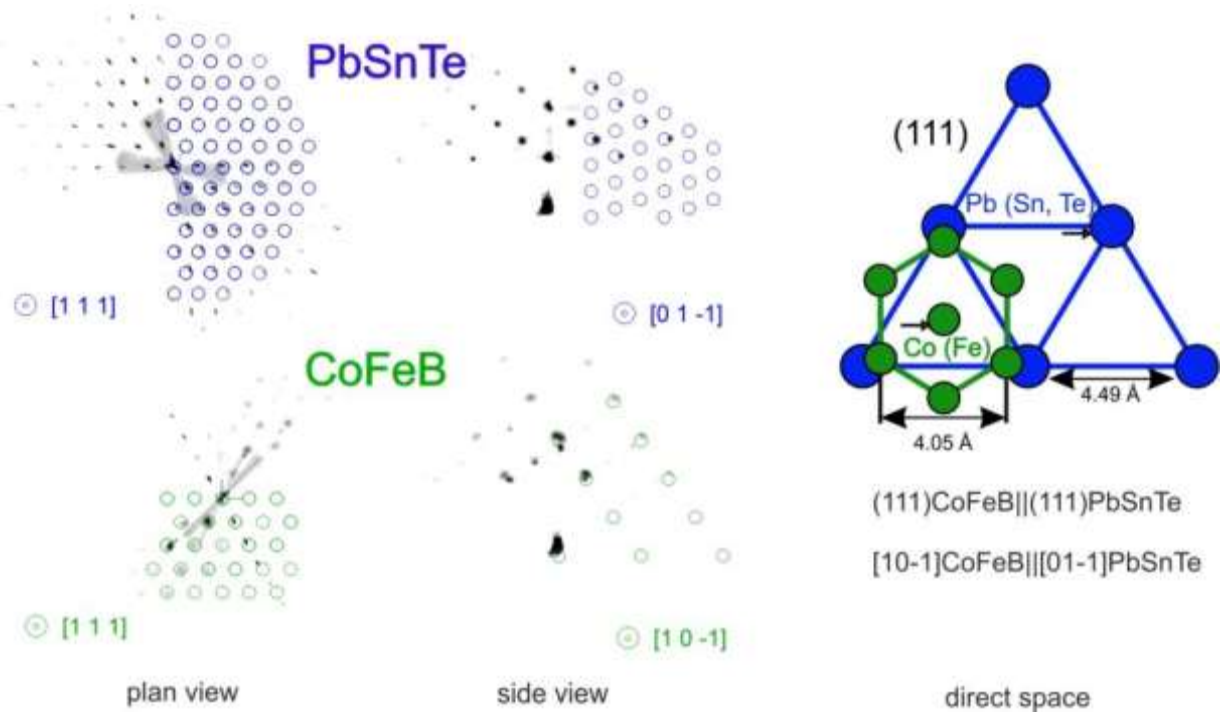

**Fig. 1.** RHEED reciprocal space maps of CoFeB(111) layer and PbSnTe (111) substrate with superimposed model reflections. RHEED maps from the surface region allow separate study of the substrate and the layer. Schematic presentation of CoFeB(111) / PbSnTe(111) in-plane lattice matching is shown on the right.

**Conclusion**

Mono-crystalline epitaxial films of ferromagnetic CoFeB on the PbSnTe topological insulator substrates were grown by the laser MBE method. The relations of main crystallographic axis of the film and the substrate were revealed using 3D reciprocal space mapping based on RHEED measurements. Quenching of some reflections of PbSnTe RHEED patterns is in agreement with the fact of the lack of Sn in compare to Pb in the substrate material, which confirms the stoichiometry of $Pb_{0.71}Sn_{0.29}Te$. The results open up further prospects for studying and using mono-crystalline ferromagnetic / topological insulator systems in the area of recording media and the systems with controllable spin-polarized current. Further studies of the magnetic and electronic properties of the ferromagnetic / topological insulator systems are underway.

*Acknowledgement*

This work has been supported by Russian Foundation of Basic Research (grants № 17-02-00729 and 17-02-00575).